\documentclass[ reprint, superscriptaddress, amsmath,amssymb, aps]{revtex4-1}
\usepackage{graphicx}
\usepackage{dcolumn}
\usepackage{amsthm}
\usepackage{bm}

\pdfoutput=1
\usepackage{graphicx,graphics,epsfig,subfigure,times,bm,bbm,amssymb,amsmath,amsthm,mathrsfs,MnSymbol}
\usepackage{gensymb}
\usepackage{amsfonts}
\usepackage[matrix,frame,arrow]{xypic}
\usepackage[pdfstartview=FitH]{hyperref}
\hypersetup{
    colorlinks=true,       
    linkcolor=red,          
    citecolor=magenta,        
    filecolor=magenta,      
    urlcolor=cyan,           
    runcolor=cyan}
\usepackage{epstopdf}
\usepackage[pdftex]{color}
\usepackage{braket}
\usepackage{enumerate}
\usepackage[normalem]{ulem}
\usepackage[usenames,dvipsnames]{xcolor}
\usepackage{multirow}
\usepackage{mathtools}

\definecolor{orange}{rgb}{1,0.5,0}

\newcommand{\ignore}[1]{}
\usepackage{geometry}

\ignore{
\documentclass[eprintnumbers,amsmath,amssymb,onecolumn,a4paper,caption ]{article}
\usepackage{amsfonts}
\usepackage{amssymb}
\usepackage{mathrsfs}
\usepackage{mathbbold}
\usepackage{bbm}
\usepackage{mathrsfs}
\usepackage{dcolumn}
\usepackage{bm}
\usepackage{times,epsfig,amssymb,amsmath}
\usepackage{float}
\usepackage{subfigure}
\usepackage{geometry}\geometry{left=2.5cm,right=2.5cm,top=3cm,bottom=3cm}

\usepackage{color}

}

\begin{document}

\title{Quantum information scrambling in the presence of weak and strong thermalization}

\author{Zheng-Hang Sun}
\affiliation{Institute of Physics, Chinese Academy of Sciences, Beijing 100190, China}
\affiliation{School of Physical Sciences, University of Chinese Academy of Sciences, Beijing 100190, China}

\author{Jian Cui}
\email{jianCui@buaa.edu.cn}
\affiliation{Department of Physics, Key Laboratory of Micro-Nano Measurement-Manipulation
and Physics (Ministry of Education), Beihang University, Beijing 100191, China}

\author{Heng Fan}
\email{hfan@iphy.ac.cn}
\affiliation{Institute of Physics, Chinese Academy of Sciences, Beijing 100190, China}
\affiliation{School of Physical Sciences, University of Chinese Academy of Sciences, Beijing 100190, China}
\affiliation{Songshan Lake  Materials Laboratory, Dongguan 523808, Guangdong, China}

\begin{abstract}
\noindent Quantum information scrambling under many-body dynamics is of fundamental interest. The tripartite mutual information can quantify the scrambling via its negative value. Here, we first study the quench dynamics of tripartite mutual information in a non-integrable Ising model where the strong and weak thermalization are observed with different initial states. We numerically show that the fastest scrambling can occur when the energy density of the chosen initial state possesses the maximum density of states. We then present an experimental protocol for observing weak and strong thermalization in a superconducting qubit array. Based on the protocol, the relation between scrambling and thermalization revealed in this work can be directly verified by superconducting quantum simulations.
\end{abstract}
\pacs{Valid PACS appear here}
\maketitle

\section{Introduction}

Under unitary dynamics, whether or not the locally encoded quantum information is delocalized and spreads over the entire system is a fundamental problem~\cite{Nature1,Nature2}. When delocalization of quantum information occurs in a system, it is referred to a scrambler. One prominent example as such is black hole, which is revealed as the most efficient scrambler~\cite{JHEP1,JHEP2,JHEP3}. The characterization of scramblers attracts considerable attention~\cite{JHEP4,JHEP_add1,JHEP5}. A well-known probe of quantum information scrambling is the out-of-time-order correlator (OTOC), whose decay rate extracted from its dynamics is closely related to the Lyapunov exponent~\cite{JHEP4,OTOC1}.
Quantum information scrambling can provide insight into the subjects in condensed-matter physics. By studying the OTOCs, it has been recognized that the scrambling plays an important role in information propagation~\cite{OTOC2,OTOC3}, many-body localization (MBL) transitions~\cite{OTOC4,OTOC5,OTOC6}, and quantum phase transitions~\cite{OTOC7,OTOC8,OTOC9,OTOC10}.

Besides the OTOC, the scrambling can also be characterized by the negative tripartite mutual information (TMI)~\cite{JHEP4,JHEP_add1}. Different from the OTOC, the TMI is an operator-independent quantity. The experimental measurements of OTOCs and TMI require different technologies. The direct measurement of OTOCs requires the inverse-time evolution, which can be performed in trapped ions~\cite{exp_OTOC1} and nuclear magnetic resonance (NMR) quantum simulators~\cite{exp_OTOC2}. Nevertheless, the inverse-time evolution is an experimental challenge in quantum superconducting circuits because of the local intra-qubit interactions~\cite{SQ_yrzhang}. On the other hand, to measure the TMI, the quantum state tomography (QST) should be employed. The accurate and efficient QST can be performed in several platforms, such as trapped ions~\cite{QST1}, superconducting qubits~\cite{SQ_yrzhang,QST2,QST3,QST_a1} and NMR~\cite{QST4}. Consequently, TMI is an experimentally feasible quantity in general.

Recently, more attention has been paid to the TMI in many-body quantum systems. It has been shown that the TMI can diagnose the ergodic and MBL phase, i.e., the TMI is close to 0 and the scrambling is suppressed in the MBL phase, while the TMI becomes smaller in the ergodic phase indicating faster scrambling~\cite{TMI1}. In addition, the scrambling is observed in both Bethe integrable system with fermionic interactions and a generic non-integrable system. In contrast, one-dimensional non-interacting fermions do not scramble~\cite{TMI2}.

Previous studies of the TMI mainly focused on its dependence on the integrability of the system. In fact, the dynamics of a quantum system is also determined by the choice of initial state. A paradigm in this regard is the weak and strong thermalization~\cite{weak_and_strong1}. Starting from an initial state corresponding to the Gibbs state with inverse temperature $\beta$ far away from 0, it has been revealed that the dynamics shows obviously persistent oscillation, which is a signature of weak thermalization. However, if one choses an initial state with $\beta\simeq 0$, strong thermalization can be observed. The weak and strong thermalization in a non-integrable Ising model with both parallel and longitude magnetic fields was numerically explored~\cite{weak_and_strong1,weak_and_strong_add1}, and explained from a quasiparticle view point~\cite{weak_and_strong2}. Recent numerical results of long-range Ising model suggest the existence of weak and strong thermalization in this system, paving the way to experimentally observing the phenomena in trapped ions~\cite{weak_and_strong3}.

In this work, we first study the time evolution of TMI in the non-integrable Ising model with different initial states, and reveal the relation between the information scrambling and the degree of thermalization. We then present an experimental protocol for observing weak and strong thermalization in a superconducting qubit array, which is one of the most popular superconducting circuit. Finally, we calculate the TMI in the superconducting qubit array, showing the relation between information scrambling and thermalization can be experimentally demonstrated on a superconducting qubit array.

\section{Results}
\subsection{Tripartite mutual information and a generic experimental protocol}
Before the calculation of TMI, three subsystems $\mathcal{A}$, $\mathcal{B}$, and $\mathcal{C}$ should be chosen, and the remainder is the subsystem $\mathcal{D}$. The reduced density matrices of the subsystems $\mathcal{A}$, $\mathcal{B}$, $\mathcal{C}$, and $\mathcal{D}$ are denoted as $\rho_{\mathcal{A}}$, $\rho_{\mathcal{B}}$, $\rho_{\mathcal{C}}$, and $\rho_{\mathcal{D}}$, respectively. The definition of TMI is~\cite{JHEP4}
\begin{eqnarray}
I_{3} = S(\rho_{\mathcal{A}}) + S(\rho_{\mathcal{B}}) + S(\rho_{\mathcal{C}}) + S(\rho_{\mathcal{D}}) \\ \nonumber
- S(\rho_{\mathcal{AB}}) - S(\rho_{\mathcal{AC}}) - S(\rho_{\mathcal{BC}}),
\label{def}
\end{eqnarray}
where $S(\rho) = -\text{Tr}(\rho\log\rho)$ is the von Neumann entropy. A negative value far away from zero is a diagnostic of quantum information scrambling~\cite{JHEP4}.

\begin{figure}
  \centering
  \includegraphics[width=1\linewidth]{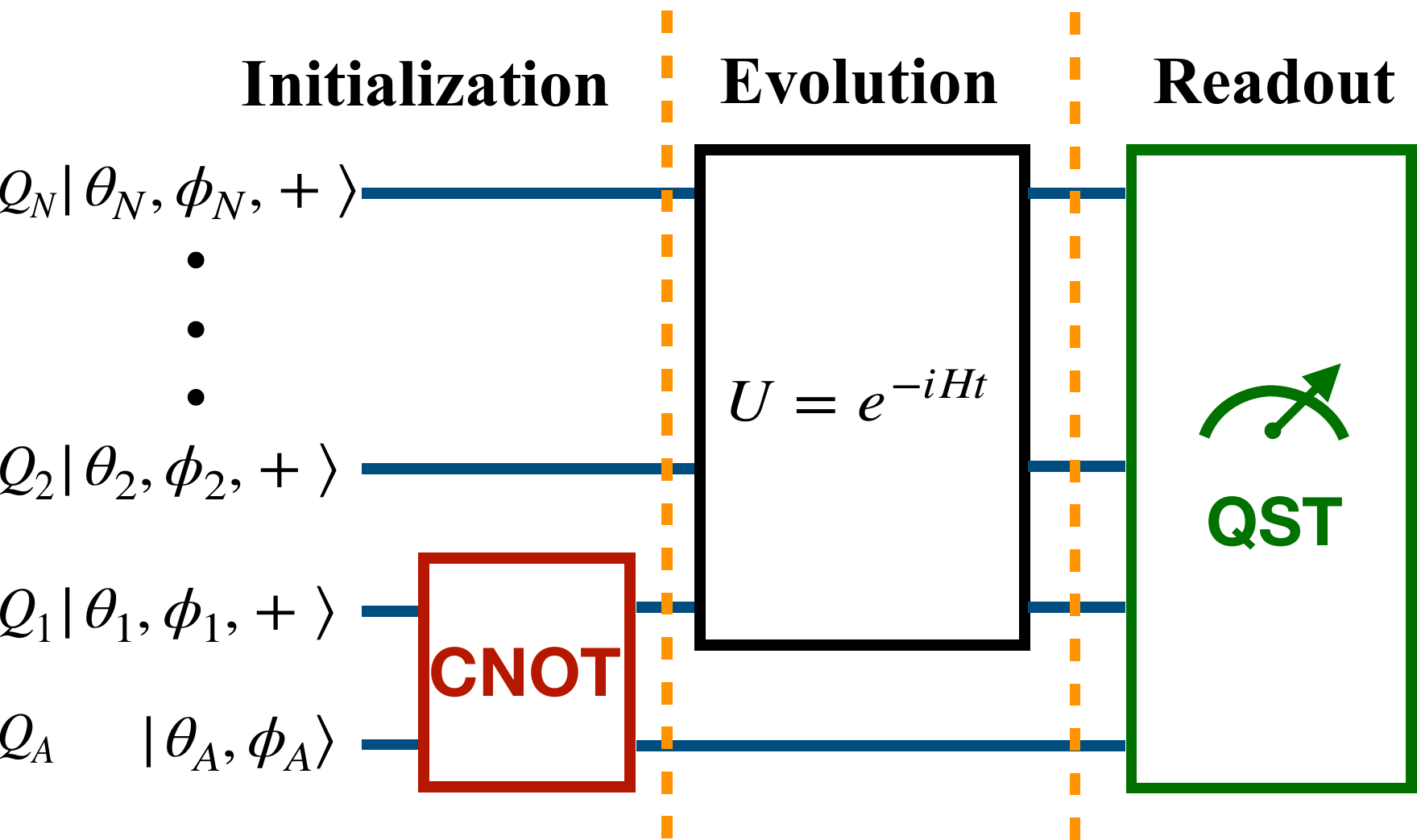}\\
  \caption{Schematic representation of the generic experimental protocol for studying quantum information scrambling by measuring the TMI. The sequence consists of three parts: (i) Initialization. (ii) Evolution. (iii) Readout. Through out the protocol, the whole system remains in pure state.}\label{sequence1}
\end{figure}

We then present a generic experimental protocol for studying quantum information scrambling by measuring the TMI. Fig.~\ref{sequence1} is a schematic diagram of the protocol. The system is comprised of $N$ qubits denoted as $Q_{1}$, $Q_{2}$, ..., and $Q_{N}$, and an ancillary qubit $Q_{A}$ (see Fig.~\ref{sequence1}). We define the state $|\theta,\phi,\pm\rangle$ as the eigenstate of the matrix $\hat{n}\cdot\vec{\sigma}=(\sin\theta\cos\phi)\sigma^{x} + (\sin\theta\sin\phi)\sigma^{y} +(\cos\theta)\sigma^{z}$ ($\sigma^{\alpha}$ with $\alpha\in\{x,y,z\}$ referring to the Pauli matrices) with the eigenvalues $\pm1$. The initial state of $Q_{A}$ is $|\theta_{A},\phi_{A}\rangle\equiv(|\theta_{1},\phi_{1},+\rangle + |\theta_{1},\phi_{1},-\rangle)/\sqrt{2}$. The  generalized $\mathrm{CNOT}$ gate in Fig.~\ref{sequence1} reads
\begin{eqnarray}
\mathrm{CNOT}\equiv|\theta_{1},\phi_{1},+\rangle\langle\theta_{1},\phi_{1},+|\otimes \mathbf{1} \\ \nonumber
 + |\theta_{1},\phi_{1},-\rangle\langle\theta_{1},\phi_{1},-|\otimes \tilde{X},
\label{CNOT}
\end{eqnarray}
where $\mathbf{1}$ is a two-dimensional identity matrix, and $\tilde{X}\equiv R \sigma^{x} R^{-1}$ with
\begin{eqnarray}
R \equiv
\begin{pmatrix}
\cos\frac{\theta_{1}}{2} & -e^{-i\phi_{1}}\sin\frac{\theta_{1}}{2} \\
e^{i\phi_{1}}\sin\frac{\theta_{1}}{2} & \cos\frac{\theta_{1}}{2}
\end{pmatrix}.
\label{R_matrix}
\end{eqnarray}
After applying the $\mathrm{CNOT}$ gate, we in fact generate a two-qubit GHZ state $|\text{GHZ}\rangle_{A1} = (|\theta_{1},\phi_{1},+\rangle_{A}|\theta_{1},\phi_{1},+\rangle_{1} + |\theta_{1},\phi_{1},-\rangle_{A}|\theta_{1},\phi_{1},-\rangle_{1})/\sqrt{2}$, entangling the ancillary qubit and $Q_{1}$ and locally encoding the information in the two qubits. In short, the initial state can be written as
\begin{eqnarray}
|\psi_{0}\rangle=|\text{GHZ}\rangle_{A1} (\otimes_{j=2}^{N}|\theta_{j},\phi_{j},+\rangle).
\label{initial}
\end{eqnarray}
Actually, when we chose $\theta_{i}=0$ or $\pi$, the aforementioned initialization is the same as the one employed in Ref.~\cite{TMI1}. This initialization is enlightened by the thought experiment for the retrieval of quantum information from a black hole~\cite{JHEP1}.

The next step is the time evolution under the quantum channel $U=e^{-iHt}$ with $H$ as the Hamiltonian of the $N$-qubits isolated system. To study the quantum information scrambling, we conventionally consider the spin chains that are beyond quadratic fermionic form after the Jordan-Wigner transformation~\cite{TMI2}. The final step is measuring the TMI by QST based on Eq. (\ref{def}). We choose  $\mathcal{A}=Q_A$,  $\mathcal{B}=Q_1$,  $\mathcal{C}=\{Q_2, Q_3,\dots, Q_{N/2}\}$, where $N$ is even for convenience, and the reminder as the subsystem  $\mathcal{D}$. The protocol can quantify how the locally encoded information scrambles through quantum dynamics.

\subsection{Results for a non-integrable Ising model}
\begin{figure*}
  \centering
  \includegraphics[width=0.8\linewidth]{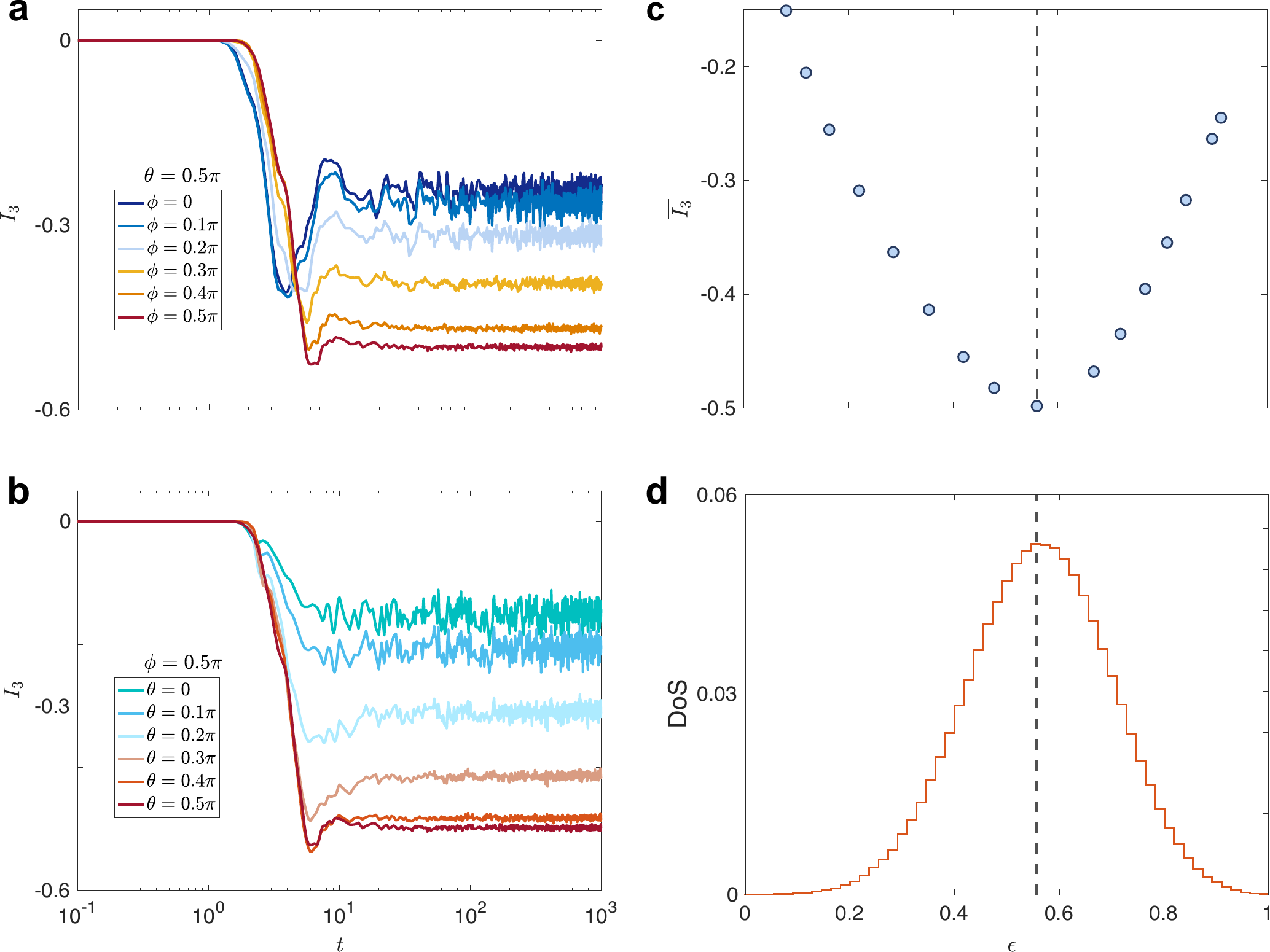}\\
  \caption{Panel \textbf{a} and \textbf{b} show the dynamics of the TMI with different isotropic initial states $|\theta,\phi\rangle$. The time-averaged TMI and the density of state (DoS) as a function of the energy density $\epsilon$ are depicted in the panel \textbf{c} and \textbf{d}, respectively. }\label{results_Ising}
\end{figure*}

We first consider a spin-1/2 Ising model whose Hamiltonian reads
\begin{eqnarray}
H_{\text{Ising}} = -J\sum_{i=1}^{N-1}\sigma_{i}^{z}\sigma_{i+1}^{z} + g\sum_{i=1}^{N}\sigma_{i}^{x} + h\sum_{i=1}^{N}\sigma_{i}^{z}
\label{Ising}
\end{eqnarray}
with $g$ and $h$ referring to the strength of the transverse and the parallel magnetic field. When $g\cdot h\neq 0$, the Ising model is non-integrable. We fix $g/J=1.05$ and $h/J=-0.5$. Similar to the quench protocol in Ref.~\cite{weak_and_strong1}, the chosen initial states are isotropic, i.e., $\theta=\theta_{i}$ and $\phi=\phi_{i}$ $\forall i\in\{1,2,...,N\}$ in Eq. (\ref{initial}), and $|\theta,\phi\rangle\equiv \otimes_{i=1}^{N}|\theta_{i},\phi_{i},+\rangle$. The inverse temperature $\beta$ for the state $|\theta,\phi\rangle$ can be obtained by solving the equation $\text{Tr}\{[\rho(\beta)-\rho(\theta,\phi)] H\}=0$ with $\rho(\theta,\phi)\equiv|\theta,\phi\rangle\langle\theta,\phi|$, $\rho(\beta)\equiv e^{-\beta H}/\text{Tr}(e^{-\beta H})$, and $H$ as the Hamiltonian.

It has been shown that the quench dynamics with the initial state $|Z+\rangle=|0,\phi,+\rangle$ (i.e., $\beta\simeq 0.7275$) shows a signature of weak thermalization, while with $|Y+\rangle=|\pi/2,\pi/2,+\rangle$ (i.e., $\beta=0$) as the initial state, strong thermalization occurs~\cite{weak_and_strong1}. Additionally, we can employ the energy density
\begin{eqnarray}
\epsilon = \frac{\langle\theta,\phi,+|H|\theta,\phi,+\rangle -E_{\text{min}}}{E_{\text{max}}-E_{\text{min}}}
\label{ed_def}
\end{eqnarray}
with $E_{\text{max(min)}}$ as the maximum (minimum) eigenvalue of the Hamiltonian $H$. The energy density can quantify the relative position of the state in the energy spectrum. It can be directly calculated that for $|Z+\rangle$ in the weak thermalization regime, the energy density $\epsilon\simeq0.0812$, i.e., $|Z+\rangle$ is quite close to the ground state of $H_{\text{Ising}}$ (Here, the system size is $N=14$), which is consistent with the quasiparticle explanation in Ref.~\cite{weak_and_strong2}. However, the energy density of $|Y+\rangle$ is $\epsilon\simeq0.5602$, far away from the ground state. We emphasize that the state $|X+\rangle=|\pi/2,0,+\rangle$ ($\beta\simeq-0.7180$ and $\epsilon\simeq0.9122$) lies in a rare region where local observables depart from their thermal values during the time evolution and no thermalization is observed~\cite{weak_and_strong1}.

Using the protocol in Fig.~\ref{sequence1}, we study the quench dynamics of TMI $I_{3}$ with several isotropic initial states parameterized by $\theta$ and $\phi$ in $H_{\text{Ising}}$ with system size $N=14$. We first demonstrate that even in the absence of thermalization, information scrambling characterized by $I_{3}<0$ can still be observed (see the result in Fig.~\ref{results_Ising}\textbf{a} with $\theta=0.5\pi$ and $\phi=0$), indicating that the occurrence of negative $I_{3}$ is independent of thermalization. Nevertheless, when $\phi$ ranges from $0$ to $0.5\pi$, the decrease of TMI suggests that stronger thermalization corresponds to faster information scrambling (see Fig.~\ref{results_Ising}\textbf{a}). Figure~\ref{results_Ising}\textbf{b} presents the dynamics of $I_{3}$ with $\phi=0.5\pi$ and $\theta \in [0, 0.5\pi]$, and shows a similar tendency of TMI in Fig.~\ref{results_Ising}\textbf{a}.

To further understand the dependence of scrambling on initial states, we consider a time-averaged TMI $\overline{I_{3}}\equiv\frac{1}{t_{f}-t_{i}}\int_{t_{i}}^{t_{f}} I_{3}(t) dt$ with $t_{i}=100$ and $t_{f}=1000$, extracting the stationary value of TMI from its dynamics. The energy density $\epsilon$ can characterize the initial states in strong or weak thermalization regime, and therefore we plot the $\overline{I_{3}}$ as a function of $\epsilon$ in Fig.~\ref{results_Ising}\textbf{c}. There is a local minimum $\overline{I_{3}}$ near $\epsilon\simeq 0.56$, which exactly corresponds to a maximum density of states (DoS) (Fig.~\ref{results_Ising}\textbf{d}), revealing that the most efficient scrambling  occurs when the initial state has the energy density with maximal DoS.

As a side remark, we compare the dynamics of TMI for different initial isotropic states with the same energy density $\epsilon$, and for the initial N\'{e}el-type and the isotropic states with the same $\epsilon$. The results are presented in Supplementary Note 1, showing that the stationary values of TMI are almost identical for all initial states with the same $\epsilon$.

\subsection{Experimental protocol for observing weak and strong thermalization on a superconducting qubit array}
\begin{figure}
  \centering
  \includegraphics[width=1\linewidth]{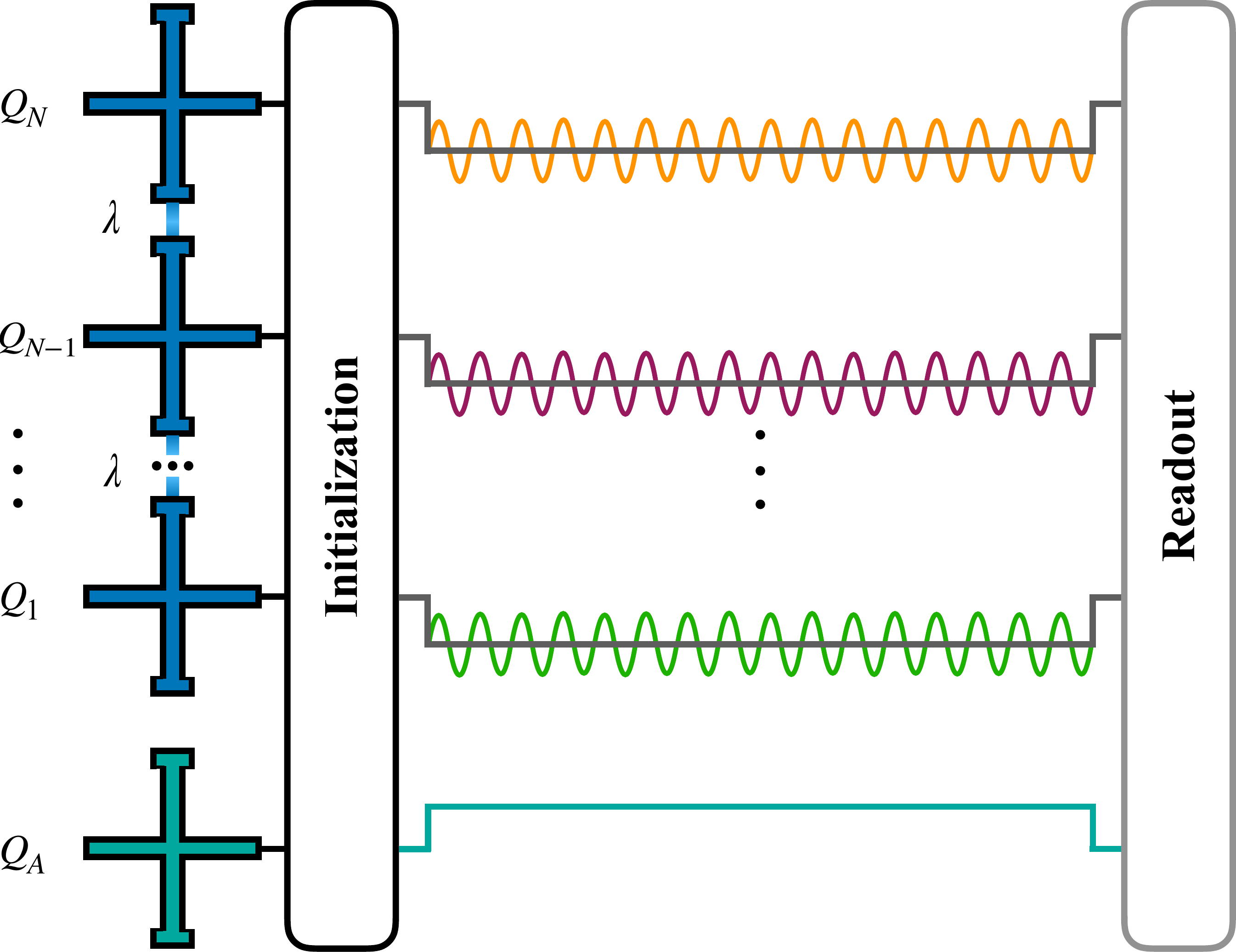}\\
  \caption{Schematic representation of the experimental waveform sequence for the time evolution. The qubit $Q_{1}$, $Q_{2}$, ..., $Q_{N}$ are biased to the interaction frequency $\omega_{q}^{\text{int.}}$ via their Z pulse control lines (the rectangular pulses). Since the ancillary qubit $Q_{A}$ does not participate in the time evolution, its frequency should be detuned away from $\omega_{q}^{\text{int.}}$. Meanwhile, the microwave drives (the sinusoidal pulses) are imposed on $Q_{1}$, $Q_{2}$, ..., $Q_{N}$ through the XY-control lines.  }\label{pulse_sequence_SQP}
\end{figure}

The one-dimensional spin-1/2 \textsl{XY} model
\begin{eqnarray}
H_{XY} = \lambda\sum_{i=1}^{N-1}(\sigma_{i}^{x}\sigma_{i+1}^{x} + \sigma_{i}^{y}\sigma_{i+1}^{y})
\label{H_0_SQP}
\end{eqnarray}
can be typically realized by a chain of transmon qubits with capacitive couplings $\lambda$~\cite{SQ_yrzhang,SQ_XY1,SQ_XY2,SQ_XY3}. The Hamiltonian $H_{XY}$ can be mapped to a free fermion system via the Jordan-Wigner transformation~\cite{free_XY}. Thus, it is well-known that thermalization is absent in $H_{XY}$ because of the infinitely many conserved quantities in thermodynamic limit (for a finite \textsl{XY} chain, there are extensive number of conserved quantities)~\cite{free_XY1,free_XY2,free_XY3}. To make the superconducting qubit array non-integrable for observing quantum thermalization, we impose uniform resonant microwave drives on all qubits, generating the local transverse field with amplitude $\Omega$, and the final Hamiltonian reads (see Methods for details)
\begin{eqnarray}
H_{\text{SQA}} = \lambda\sum_{i=1}^{N-1}(\sigma_{i}^{x}\sigma_{i+1}^{x} + \sigma_{i}^{y}\sigma_{i+1}^{y}) + \Omega\sum_{i=1}^{N} \sigma_{i}^{y}.
\label{H_1_SQP}
\end{eqnarray}
The local transverse field has been realized in a recent quantum simulation experiment~\cite{DQPT}, where the $XY$-crosstalk correction and phase alignment of the transverse field were discussed. A sketch of the pulse sequence for the realization of the Hamiltonian (\ref{H_1_SQP}) is depicted in Fig.~\ref{pulse_sequence_SQP}.

Before we study the TMI in the Hamiltonian (\ref{H_1_SQP}), the weak and strong thermalization in the system should be demonstrated. Here, we adopt $\lambda=\Omega=1$ and $N=14$. It can be calculated that for the isotropic initial state $|\theta,\phi\rangle=|\pi/2,1.369\pi\rangle$, the inverse temperature $\beta\simeq 0$, and the strong thermalization is expected. To observe the weak thermalization, we consider another isotropic initial state $|\pi/2,0.369\pi\rangle$ with $\beta\simeq -0.6547$. Moreover, we recognize that the energy of a typical initial state $|Z+\rangle$ is $\langle Z+|H_{\text{SQA}}|Z+\rangle = 0$. Thus, for $|Z+\rangle$, $\beta=0$, and $|Z+\rangle$ lies in the strong thermalization regime. In supplementary Note 2, we show that the dynamical properties of $H_{\text{SQA}}$ with the initial state $|Z+\rangle$ are similar to those with the initial state $|\pi/2,1.369\rangle$.

\begin{figure*}
  \centering
  \includegraphics[width=0.9\linewidth]{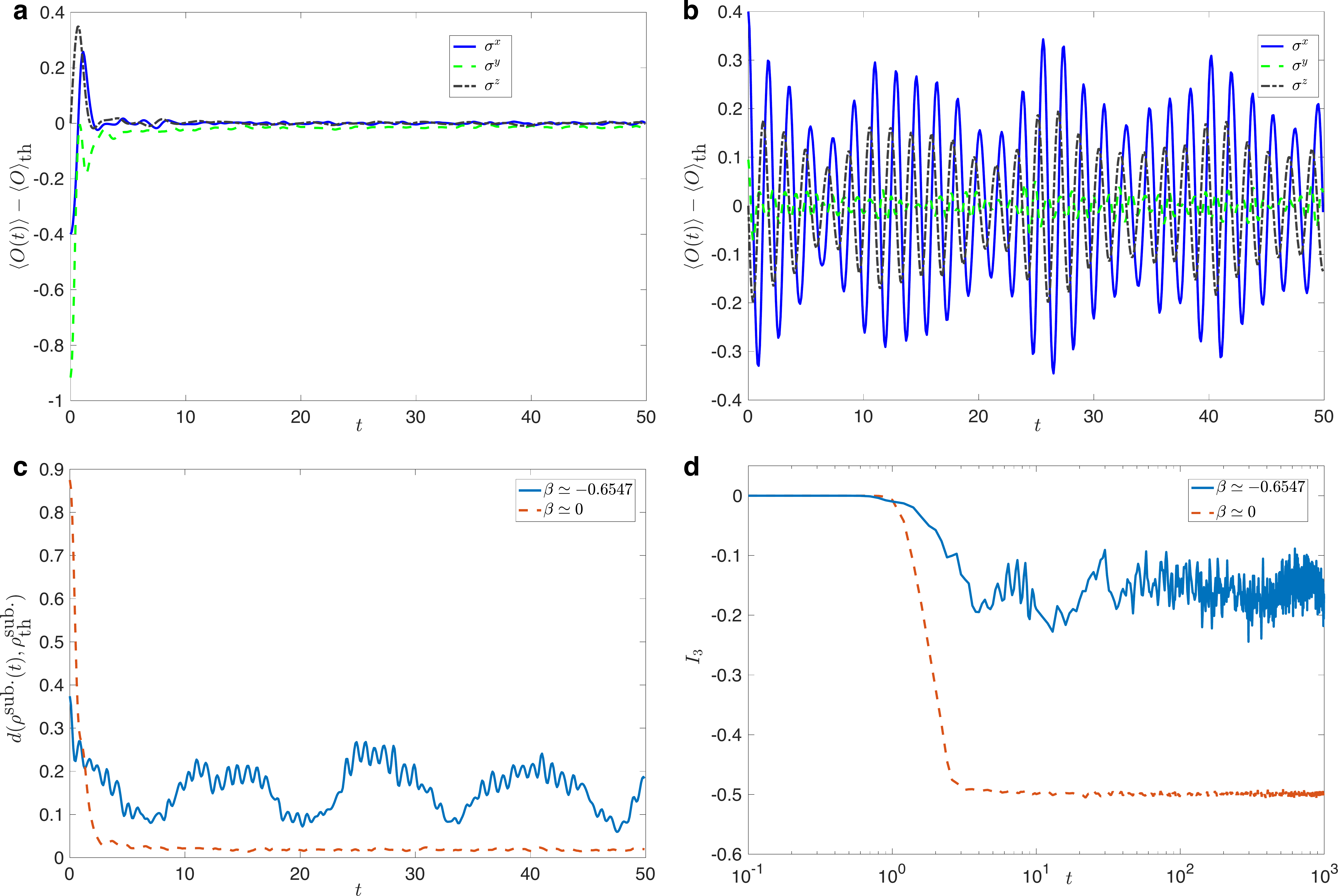}\\
  \caption{Panel \textbf{a} and \textbf{b} show the dynamics of local observables with the initial state $|\theta,\phi\rangle=|\pi/2,1.369\pi\rangle$ ($\beta\simeq 0$) and $|\theta,\phi\rangle=|\pi/2,0.369\pi\rangle$ ($\beta\simeq-0.6547$), respectively. Panel \textbf{c} shows the distance $d(\rho^{\text{sub.}}(t),\rho^{\text{sub.}}_{\text{th}})$ with the two initial states. Panel \textbf{d} shows the dynamics of the TMI $I_{3}$ with the two initial states.}\label{weak_and_strong_SQP}
\end{figure*}

Similar to Ref.~\cite{weak_and_strong1}, we pay attention to the quench dynamics of local observables $\langle O(t)\rangle - \langle O\rangle_{\text{th}}$ with $O\in\{\sigma^{x},\sigma^{y},\sigma^{z}\}$, and the operator norm distance between a reduced density matrix (RDM) of a three-body subsystem and the corresponding thermal density matrix denoted as $d(\rho^{\text{sub.}}(t),\rho^{\text{sub.}}_{\text{th}})$ (see Methods for the detailed definitions). The time evolution of local observables with two different initial states are shown in Fig.~\ref{weak_and_strong_SQP}\textbf{a} and \textbf{b}. One can see that the values of the observables relax to the thermal values when the initial state is $|\theta,\phi\rangle=|\pi/2,1.369\pi\rangle$ ($\beta\simeq 0$) and in the strong thermalization region. However, the undamped oscillation of $\langle O(t)\rangle - \langle O\rangle_{\text{th}}$ can be observed with another initial state $|\theta,\phi\rangle=|\pi/2,0.369\pi\rangle$, which is a signature of the weak thermalization. Moreover, figure~\ref{weak_and_strong_SQP}\textbf{c} shows the dynamics of $d(\rho^{\text{sub.}}(t),\rho^{\text{sub.}}_{\text{th}})$ with the two initial states. In the strong thermalization region, the quenched RDM fast saturates to the thermal state and the distance $d(\rho^{\text{sub.}}(t),\rho^{\text{sub.}}_{\text{th}})$ monotonically decays, while the distance exhibits dramatic fluctuation in the weak thermalization region.

\subsection{Results for the superconducting qubit array}
\begin{figure}
  \centering
  \includegraphics[width=0.9\linewidth]{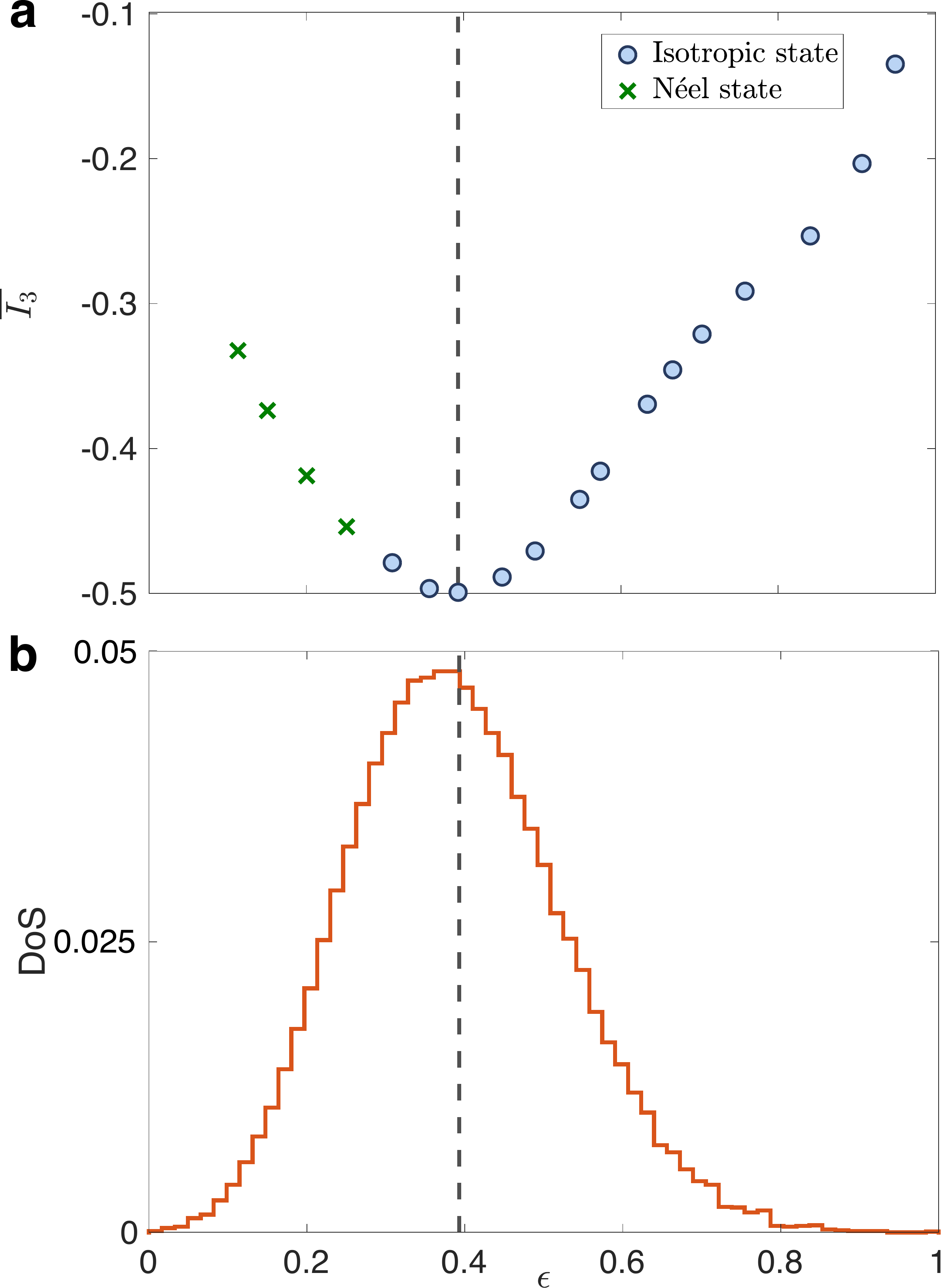}\\
  \caption{For the Hamiltonian (\ref{H_1_SQP}), the time-averaged TMI and the density of state (DoS) as a function of the energy density $\epsilon$ are depicted in the panel \textbf{a} and \textbf{b}, respectively. }\label{TMI_SQP}
\end{figure}

We then study the TMI $I_{3}$ in the Hamiltonian (\ref{H_1_SQP}). Figure ~\ref{weak_and_strong_SQP}\textbf{d} presents the time evolution of $I_{3}$ with the two initial states. The behaviors of $I_{3}$ are similar to those in the Ising model (see Fig.~\ref{results_Ising}\textbf{a} and \textbf{d}). Fast and slow quantum information scrambling are observed in the strong and weak thermalization region, respectively. Moreover, the saturation of $I_{3}$ at long time can also be observed in the system (\ref{H_1_SQP}).

Next, we study the relation between the TMI and the energy density $\epsilon$. Different from the Ising model (\ref{Ising}), the minimum attainable $\epsilon$ of all isotropic states is $0.3093$ in the system (\ref{H_1_SQP}) (see Supplementary Information Note 1). To study the $I_{3}$ of the initial states with $\epsilon<0.3093$, we can consider the N\'{e}el-type initial states (see Supplementary Information Note 1). Figure~\ref{TMI_SQP}\textbf{a} depicts the time-averaged TMI  $\overline{I_{3}}\equiv\frac{1}{t_{f}-t_{i}}\int_{t_{i}}^{t_{f}} I_{3}(t) dt$ ($t_{i}=100$ and $t_{f}=1000$) as a function of $\epsilon$. We also present the DoS of the Hamiltonian (\ref{H_1_SQP}) as a function of $\epsilon$ in Fig.~\ref{TMI_SQP}\textbf{b}. The comparison between Fig.~\ref{TMI_SQP}\textbf{a} and \textbf{b} indicates that the fastest information scrambling occurs when the initial state has the $\epsilon$ with maximum DoS. The results of $I_{3}$ in the Hamiltonian (\ref{H_1_SQP}) suggest that the linkage between information scrambling and thermalization revealed in the Ising model (\ref{Ising}) can be experimentally verified in the superconducting qubit array using the aforementioned protocol.


%
%

\section{Discussion}
We have investigated the quantum information scrambling in the systems where the strong and weak thermalization exist. The occurrence of information scrambling quantified by the negative TMI, i.e., $I_{3}<0$, can be observed in both strong and weak thermalization regimes. Ref.~\cite{weak_and_strong1} indicates the absence of quantum thermalization for the initial state $|X+\rangle$ in the Ising model (\ref{Ising}). However, our result in figure~\ref{results_Ising}\textbf{a} shows information scrambling during its quench dynamics with the initial state $|X+\rangle$. Actually, it has been pointed out that scrambling does not equal quantum thermalization~\cite{TMI1,OTOC_add1}, and our results are consistent with the viewpoint. Nevertheless, the existence of no thermalization region in the Ising model remains an open question due to the two numerical challenges. Firstly, the matrix product state (MPS) algorithm can only simulate short-time evolution~\cite{weak_and_strong1}, while one has to compute long-time dynamics reaching thermalization time to obtain more solid evidence. Secondly, a recent numerical work using MPS combined with Chebyshev polynomial expansions shows there are significantly large error bars of the results for the states in the no thermalization region since the states are close to the edge of the DoS spectrum~\cite{weak_and_strong_add_MC}. In Supplementary Note3, we present a finite-size analysis of the no thermalization region employing the Lanczos-Krylov space-time calculations, suggesting that the results of the infinite system in Ref.~\cite{weak_and_strong1} support the finite-size results.

We reveal that the value of $I_{3}$ is closely related to the weak and strong thermalization characterized via the energy density and DoS. Ref.~\cite{weak_and_strong3} reveals the weak and strong thermalization in the long-range Ising model, which can be realized in tapped-ion quantum simulator~\cite{TI_sum}. As a consequence, this work may inspire the investigations on the quantum information scrambling in long-range interacting systems~\cite{long_range_TMI}, especially in the presence of weak and strong thermalization~\cite{weak_and_strong3}. Besides the weak thermalization, the weak ergodicity breaking characterized by the long-lived oscillations can also be originated from quantum many-body scars~\cite{scar1,scar2}, and the study of information scrambling with quantum many-body scars is another intriguing direction to explore in the future.

Quantum thermalization dynamics has been experimentally studied in optical lattice~\cite{exp_thermal_1} and tapped ions~\cite{exp_thermal_2}. Previous experiments mainly focus on the strong thermalization (which is actually regarded as the conventional quantum thermalization), and a distinct comparison between weak and strong thermalization remains absent. We develops a scheme to probe weak and strong thermalization in a superconducting qubit array, and the results can be readily demonstrated in quantum simulation experiments.

It has been shown that the energy density plays a key role in the MBL mobility edge~\cite{MBME_1,MBME_2,MBME_3,MBME_4}. The relation between information scrambling and energy density revealed by our work could provide new insight into the mobility edge. Furthermore, the slow information scrambling in the weak thermalization region has potential applications for quantum information storage devices~\cite{QST_a1} and stabilizing out-of-equilibrium phases of matter~\cite{non_eq_phase1,non_eq_phase2}.

\section{Methods}
\subsection{Realization of the Hamiltonian (\ref{H_1_SQP}) in a superconducting qubit array}
Conventionally, the Hamiltonian of a transmon qubit array can be described by the Bose-Hubbard model~\cite{SQ_yrzhang,QST_a1}
\begin{eqnarray}
H_{\text{BH}} = \Lambda\sum_{i=1}^{N-1}(a_{i}^{\dagger}a_{i+1} + a_{i}a_{i+1}^{\dagger}) + \frac{U}{2}\sum_{i=1}^{N} n_{i}(n_{i} - 1)
\label{H_BH}
\end{eqnarray}
with $a_{i}^{\dagger}$ ($a_{i}$) as the bosonic creation (annihilation) operator, $n_{i}=a_{i}^{\dagger}a_{i}$, $U$ as strength of nonlinear interaction and $\Lambda$ referring to the nearest hopping strength. In the limit $U/\Lambda\rightarrow\infty$, the Hamiltonian~(\ref{H_BH}) reduces to a $XY$ model~\cite{BH1,BH2}
\begin{eqnarray}
\label{H_XY_add}
H_{XY} &=& \Lambda\sum_{i=1}^{N-1}(\sigma_{i}^{+}\sigma_{i+1}^{-} + \sigma_{i}^{-}\sigma_{i+1}^{+}) \\ \nonumber
&=& \lambda\sum_{i=1}^{N-1}(\sigma_{i}^{x}\sigma_{i+1}^{x} + \sigma_{i}^{y}\sigma_{i+1}^{y})
\end{eqnarray}
with $\lambda = \Lambda/2$.

Actually, the non-equilibrium properties of the Bose-Hubbard model~(\ref{H_BH}) are close to those of the $XY$ model~(\ref{H_XY_add}) when $U/\Lambda\geq 8$~\cite{BH2}. For the device in Ref.~\cite{SQ_yrzhang}, $U/\Lambda\simeq 18$, and the $XY$ model can be experimentally studied using analog quantum simulation.

When the microwave drives with amplitude $\Omega$ are applied to each qubit, we can obtain~\cite{drive_add}
\begin{eqnarray}
H_{\text{drive}} = \Omega\sum_{j=1}^{N} e^{-i\varphi_{j}}\sigma_{j}^{+} + e^{i\varphi_{j}}\sigma_{j}^{-}.
\label{H_drive}
\end{eqnarray}
By adjusting the phase of the microwave drives, one can force $\varphi=\varphi_{j}=\pi/2$ ($j=1,2,...,N$), and then $H_{\text{drive}}$ can be rewritten as $H_{\text{drive}}=\Omega\sum_{j=1}^{N}\sigma_{j}^{y}$. Thus, the Hamiltonian~(\ref{H_1_SQP}), i.e., $H_{\text{SQA}} = H_{XY} + H_{\text{drive}}$ can be realized in a qubit array.

To better understand the strong thermalization in the system~(\ref{H_1_SQP}), we can rewrite the Hamiltonian in the $\sigma^{x}$ basis
\begin{eqnarray}
\label{H_SQP_sigmax}
H_{\text{SQP}} &=& \Lambda\sum_{i=1}^{N-1} \sigma_{i}^{z}\sigma_{i+1}^{z} + \Omega\sum_{i=1}^{N}\sigma_{i}^{x} + \Omega\sum_{i=1}^{N-1}\sigma_{i}^{x}\sigma_{i+1}^{x} \\ \nonumber
&=& H_{0} + H_{\text{int.}}.
\end{eqnarray}
By employing the Jordan-Wigner transformation $\sigma_{i}^{x}=1-2c^{\dagger}_{i}c_{i}$ and $\sigma_{i}^{z} = -\prod_{l<i}(1-2c^{\dagger}_{l}c_{l})(c_{i} + c_{i}^{\dagger})$ with $c_{i}^{\dagger}$ ($c_{i}$) referring to the fermionic creation (annihilation) operator, one can see that in the Hamiltonian~(\ref{H_SQP_sigmax}), $H_{0} \equiv \Lambda\sum_{i=1}^{N-1} \sigma_{i}^{z}\sigma_{i+1}^{z}+ \Omega\sum_{i=1}^{N}\sigma_{i}^{x}$ as the Ising model without parallel field can be regarded as a quadratic system (free fermions). Moreover, the $H_{\text{int.}}\equiv \Omega\sum_{i=1}^{N-1}\sigma_{i}^{x}\sigma_{i+1}^{x}$ gives the Heisenberg coupling $c_{i}c_{i}c_{i+1}c_{i+1}$, from which the quantum thermalization and MBL are originated~\cite{MBL_add1}. Based on above discussions, we explain the occurrence of the strong thermalization in the superconducting qubit array.

\subsection{The quantities employed to quantify the quantum thermalization}
Here, we briefly introduce the definitions of the quantities that quantify the quantum thermalization. During the quench dynamics, there exists an exchange of information from a small subsystem $\mathcal{A}$ to the complementary one $\overline{\mathcal{A}}$ that acts as a thermal bath of $\mathcal{A}$. The reduced density operator of $\mathcal{A}$ at time $t$ is $\rho^{\mathcal{A}}(t)\equiv \text{Tr}_{\overline{\mathcal{A}}}\{\rho(t)\}$ with $\rho(t)$ as the quenched state. The thermal density matrix of the same system equilibrium at temperature $T$ is $\rho_{\text{th.}}(T)=Z^{-1}\exp(-\beta H)$ with $\beta = 1/T$ as the inverse temperature and $Z\equiv \text{Tr}\{\exp(-\beta H)\}$, and thus $\rho^{\mathcal{A}}_{\text{th.}}\equiv \text{Tr}_{\overline{\mathcal{A}}}\{\rho_{\text{th.}}(T)\}$. When the quantum thermalization occurs, with a long time $t$ and a temperature $T$, $\rho^{\mathcal{A}}(t) = \rho_{\text{th.}}^{\mathcal{A}}(T)$~\cite{thermalization}.

As a direct consequence, if a quantity can measure the distance between $\rho^{\mathcal{A}}(t)$ and $\rho_{\text{th.}}^{\mathcal{A}}(T)$, it can be employed to quantify the quantum thermalization. In Ref.~\cite{weak_and_strong1} and our work, two quantities are considered. The first one is related to the local observables, whose definition is
\begin{eqnarray}
\langle\mathcal{O}(t)\rangle - \langle\mathcal{O}\rangle_{\text{th.}} = \text{Tr}\{\mathcal{O}(\rho^{\mathcal{A}}(t) - \rho_{\text{th.}}^{\mathcal{A}}(T))\},
\label{local_observable}
\end{eqnarray}
where $\mathcal{O}$ is the local observables $\sum_{i=1}^{N}\sigma_{i}^{\alpha}/N$ ($\alpha\in\{x,y,z\}$). The second one is the distance $d(\rho^{\mathcal{A}}(t),\rho_{\text{th.}}^{\mathcal{A}}(T))$ defined as the maximum eigenvalue of the matrix $\rho^{\mathcal{A}}(t)-\rho_{\text{th.}}^{\mathcal{A}}(T)$. In this work, we chose the subsystem $\mathcal{A}$ consisting of the qubit $Q_{5}$, $Q_{6}$ and $Q_{7}$.

\begin{acknowledgments}
We acknowledge the enlightening discussion with M. C. Ba\~{n}uls. This work was supported by NSFC (Grant Nos.~11904018, 11774406, 11934018), National Key R$\&$D Program of China (Grant Nos.~2016YFA0302104, 2016YFA0300600), Strategic Priority Research Program of Chinese Academy of Sciences (Grant No.~XDB28000000), and Beijing Academy of Quantum Information (Grant No.~Y18G07).
\end{acknowledgments}

%

\clearpage

\noindent \textbf{Supplementary Note 1. The dynamics of tripartite mutual information for different initial states with identical energy density}

\begin{figure*}[h]
  \centering
  \includegraphics[width=1\linewidth]{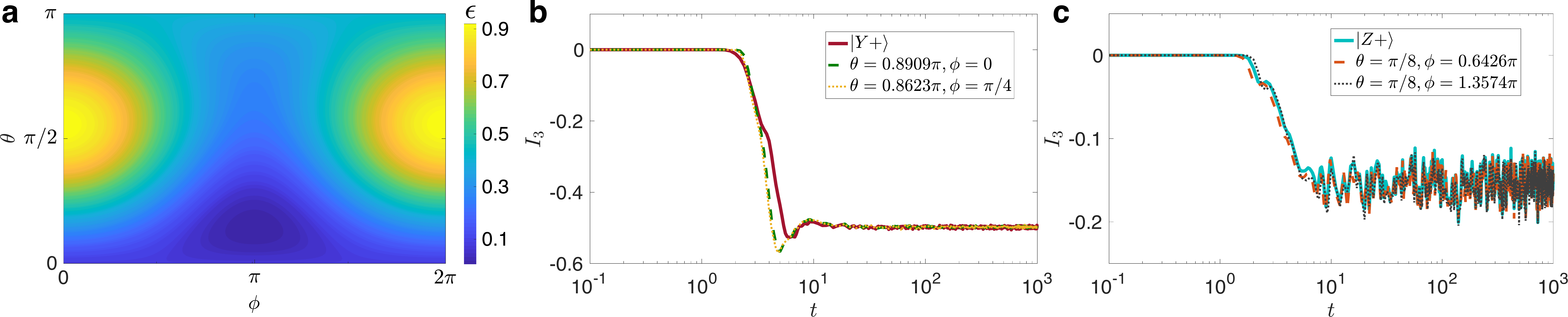}\\
  \caption{\textbf{a} The energy density $\epsilon$ as a function of $\theta$ and $\phi$ in the non-integrable Ising model studied in the main text. \textbf{b} The quench dynamics of TMI with different isotropic initial states with the same energy density $\epsilon = 0.5602$. \textbf{c} is similar to \textbf{b} but with another  energy density $\epsilon = 0.0812$.}\label{S1}
\end{figure*}

We present several additional results supporting the main text. We first plot the energy density of isotropic initial states $|\theta,\phi,+\rangle$, defined as $\epsilon(\theta,\phi)\equiv (\langle\theta,\phi,+|H_{\text{Ising}}|\theta,\phi,+\rangle - E_{\text{min}}) /(E_{\text{max}} - E_{\text{min}})$ in the $\theta$-$\phi$ plane in Fig.~\ref{S1}\textbf{a}. We then benchmark the dynamics of tripartite mutual information (TMI) for different initial states $|\theta,\phi,+\rangle$ with the same randomly chosen $\epsilon$, and the results are shown in Fig.~\ref{S1}\textbf{b} and \textbf{c}. It is seen that the saturated values of TMI are approximately equal to each other, if the energy densities are identical.

\begin{figure*}[h]
  \centering
  \includegraphics[width=0.8\linewidth]{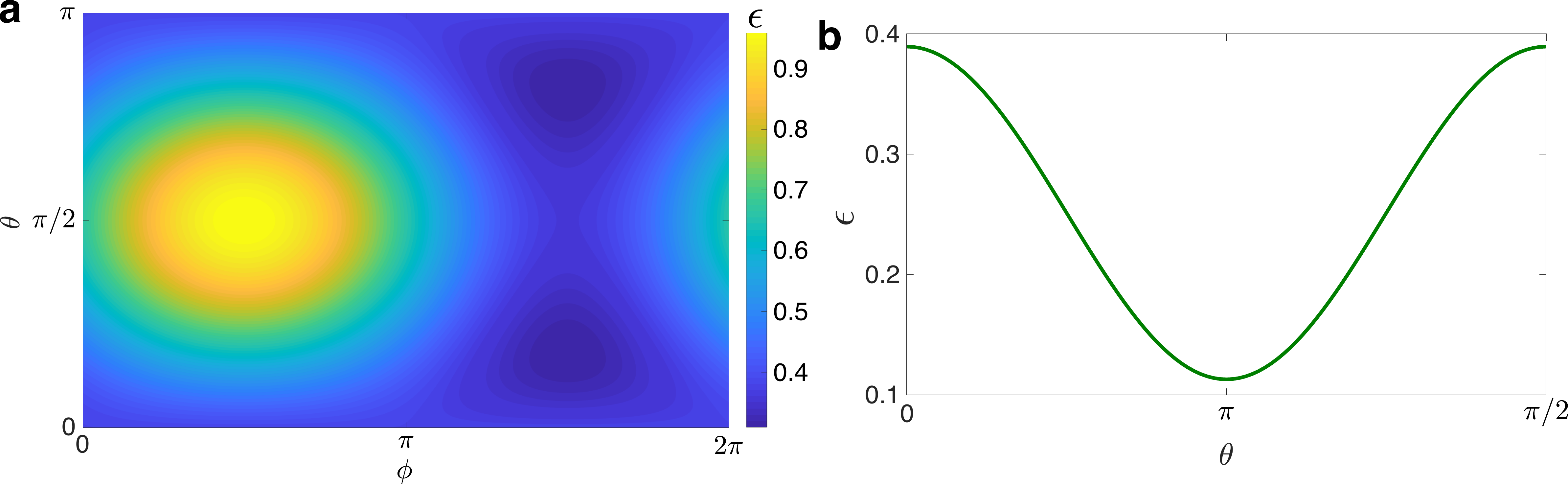}\\
  \caption{\textbf{a} The energy density $\epsilon$ of isotropic initial states as a function of $\theta$ and $\phi$ in the superconducting qubit array studied in the main text. \textbf{b} The energy density $\epsilon$ of N\'{e}el-type initial states as a function of $\theta$ with $\phi=0$.}\label{S2}
\end{figure*}

Next, we calculate the energy density of $|\theta,\phi,+\rangle$ in the superconducting qubit array, i.e., the Hamiltonian (8) in the main text. The results are depicted in Fig.~\ref{S2}\textbf{a}, showing that the energy density $\epsilon<0.3093$ are not attainable for isotropic initial states. Thus, we consider another type of initial states, i.e., the N\'{e}el-type initial states
\begin{eqnarray}
|\psi_{0}\rangle = |\text{GHZ}\rangle_{A1} \otimes |\theta_{2},\phi_{2},-\rangle \otimes |\theta_{3},\phi_{3},+\rangle \otimes ... \otimes |\theta_{N-1},\phi_{N-1},+\rangle \otimes |\theta_{N},\phi_{N},-\rangle
\label{neel}
\end{eqnarray}
with $\theta_{i} = \theta$, $\phi_{i}=\phi$ ($i=1,2,...,N$) and the number of qubits $N$ as a even number for convenience. We further calculate the energy density of the N\'{e}el-type initial states with $\phi=0$ and several values of $\theta$. As shown in Fig.~\ref{S2}\textbf{b}, the dynamics of TMI for the initial states with $\epsilon<0.3093$ are available by choosing the N\'{e}el-type initial states.

\begin{figure*}[h]
  \centering
  \includegraphics[width=0.8\linewidth]{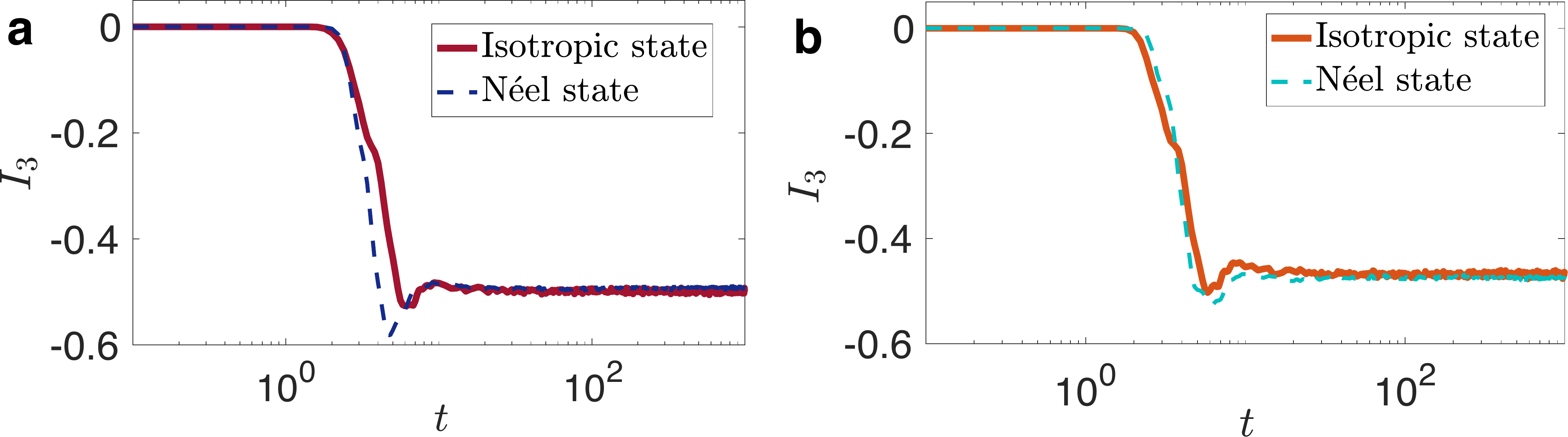}\\
  \caption{\textbf{a} In the non-integrable Ising model, the dynamics of TMI $I_{3}$ for the isotropic initial state $|Y+\rangle$ and the N\'{e}el-type initial state with $\theta=\pi/2$. The value of energy density for both initial states is $\epsilon=0.5602$.  \textbf{b} The dynamics of $I_{3}$ for the isotropic initial state $|\theta=0.5\pi,\phi=0.4\pi\rangle$ and the N\'{e}el-type initial state with $\theta=0.7013\pi$. The value of energy density for both initial states is $\epsilon=0.6690$. }\label{S3}
\end{figure*}

\begin{figure*}[h]
  \centering
  \includegraphics[width=0.8\linewidth]{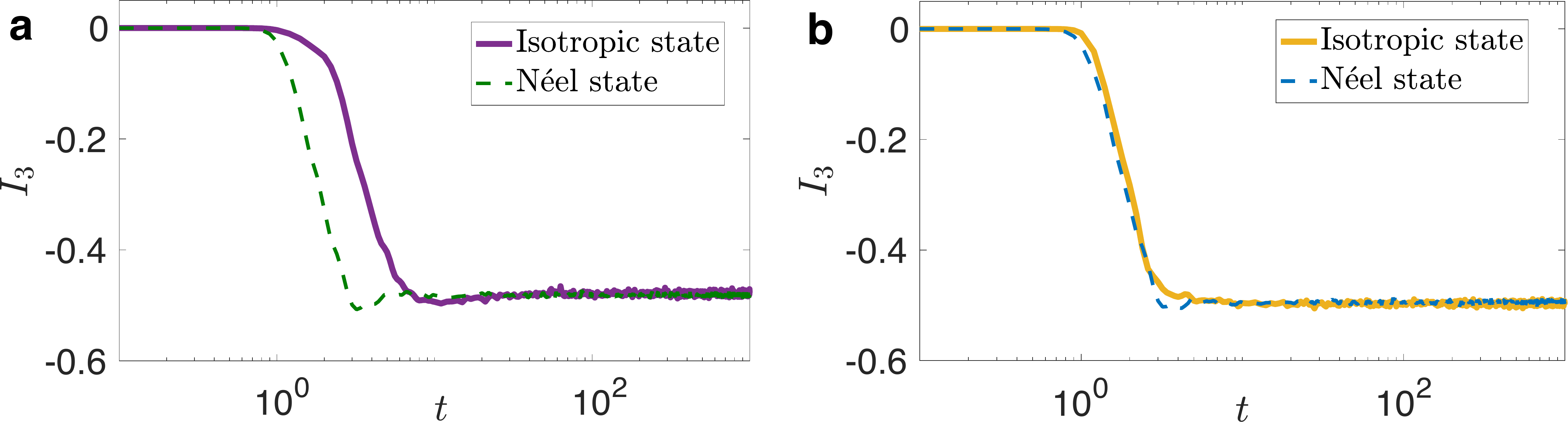}\\
  \caption{\textbf{a} In the superconducting qubit array, the dynamics of TMI $I_{3}$ for the isotropic initial state $|\theta=0.18\pi,\phi=1.5\pi\rangle$ and the N\'{e}el-type initial state with $\theta=0.1810\pi$. The value of energy density for both initial states is $\epsilon=0.3093$.  \textbf{b} The dynamics of $I_{3}$ for the isotropic initial state $|\theta=0.4\pi,\phi=1.5\pi\rangle$ and the N\'{e}el-type initial state with $\theta=0.1124\pi$. The value of energy density for both initial states is $\epsilon=0.3564$.}\label{S4}
\end{figure*}

Figure~\ref{S1}\textbf{b} and \text{c} suggest that for the isotropic state, the saturated value of TMI is directly dependent on the $\epsilon$. Therefore, it can be predicted that the saturated value of TMI is insensitive to the specific initial states, i.e., the isotropic or N\'{e}el-type states, if the $\epsilon$ of the states are equal to each other. The results in Fig.~\ref{S3} and~\ref{S4} provide further evidence for this prediction.

\clearpage

\noindent \textbf{Supplementary Note 2. Non-equilibrium properties of the superconducting qubit array with the initial state $|Z+\rangle$}

\begin{figure*}[h]
  \centering
  \includegraphics[width=1\linewidth]{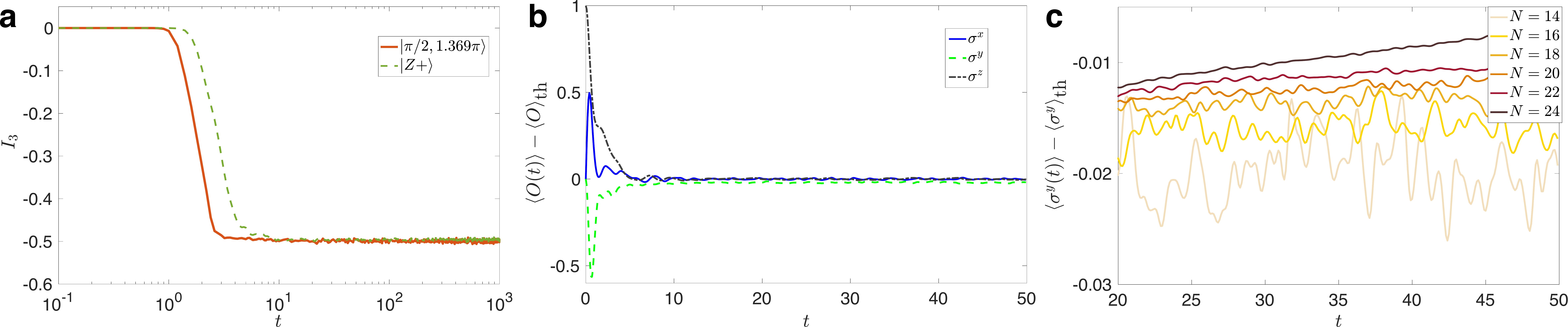}\\
  \caption{\textbf{a} The dynamics of TMI with the initial state $|\pi/2,1.369\pi\rangle$ and $|Z+\rangle$, whose stationary value is close to each other.   \textbf{b} The dynamics of local observables with $N=14$ and the initial state $|Z+\rangle$. \textbf{c} The dynamics of a local observable $\langle\sigma^{y}(t)\rangle - \langle\sigma^{y}\rangle_{\text{th}}$ with different $N$.}\label{S5}
\end{figure*}

It can be directly calculated that for the superconducting qubit array considered in the main text, the inverse temperature of the initial state $|Z+\rangle$ is $\beta=0$. Hence, $|Z+\rangle$ lies in the strong thermalization regime. In the main text, we have shown that the $\beta$ of initial state $|\pi/2,1.369\pi\rangle$ is also equal to 0. Here, we present a comparison between the dynamics of TMI with the initial state $|\pi/2,1.369\pi\rangle$ and $|Z+\rangle$ in Fig.~\ref{S5}\textbf{a}, which is consistent with the results in Fig.~\ref{S1}, \ref{S3} and \ref{S4}.

Moreover, we depict the time evolution of local observables with the initial state $|Z+\rangle$ in Fig.~\ref{S5}\textbf{b}. The dynamical behaviors of local observables are similar to those in the Fig. 4\textbf{a} of the main text. With the number of qubit $N=14$, $\langle\sigma^{\alpha}(t)\rangle - \langle\sigma^{\alpha}\rangle_{\text{th}}$ ($\alpha\in\{x,z\}$) fast tend to 0 after relaxation. However, $\langle\sigma^{y}(t)\rangle - \langle\sigma^{y}\rangle_{\text{th}}$ suffers from a stronger finite-size effect. We then present the results of $\langle\sigma^{y}(t)\rangle - \langle\sigma^{y}\rangle_{\text{th}}$ with larger $N$ in Fig.~\ref{S5}\textbf{c}. With the increase of $N$, the value of $\langle\sigma^{y}(t)\rangle - \langle\sigma^{y}\rangle_{\text{th}}$ becomes closer to 0.

\noindent \textbf{Supplementary Note 3. A finite-size analysis of the no thermalization regime}

The no thermalization regime in the non-integrable Ising model has been numerically revealed using the matrix product state (MPS) algorithm that can simulate infinite quantum systems. A characteristic of the no thermalization regime is that the dynamics of local observables depart from their thermal values. For instance, in the non-integrable Ising model, with the initial state $|X+\rangle$, the dynamics of the local observable $\langle\sigma^{x}(t)\rangle$ does not converge to the thermal value $\langle\sigma^{x}\rangle_{\text{th}}$ (please see Ref. [29] and [30] for more details).

\begin{figure*}[h]
  \centering
  \includegraphics[width=0.9\linewidth]{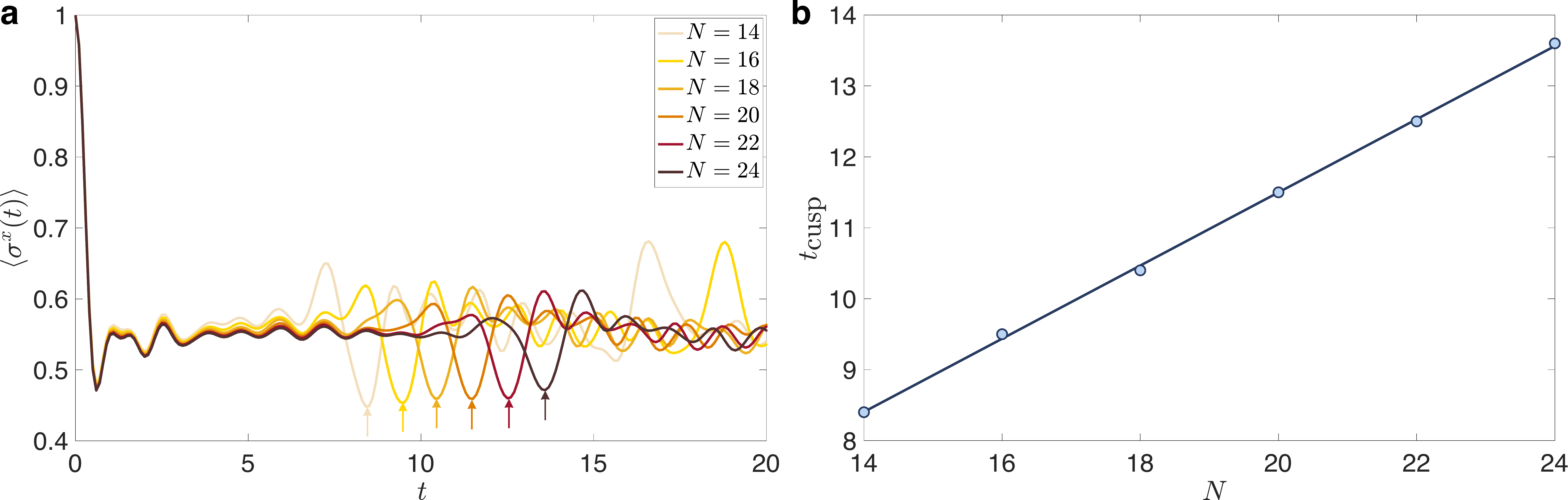}\\
  \caption{\textbf{a} The dynamics of $\langle\sigma^{x}(t)\rangle$ in the non-integrable Ising model with the initial state $|X+\rangle$ and different system size $N$. \textbf{b} The dependence of $t_{\text{cusp}}$ and $N$.  }\label{S6}
\end{figure*}

Here, we present the numerical results of the local observable $\langle\sigma^{x}(t)\rangle$ with finite system size $N$ in the non-integrable Ising model in Fig.~\ref{S6}\textbf{a}. It is seen that the signature of no thermalization regime suffers from strong finite size effect. The approximately monotonous increase of $\langle\sigma^{x}(t)\rangle$ after the relaxation observed in the infinite system is interrupted by the obvious oscillations marked by the arrows in Fig.~\ref{S6}\textbf{a}. We recognize that the oscillations are dependent on the system size $N$. As a reasonable shortcut, we can study the location of the first dramatic cusp $t_{\text{cusp}}$ marked by the arrows in Fig.~\ref{S6}\textbf{a} as a function of $N$, and the results are presented in Fig.~\ref{S6}\textbf{b}. Up to the system size $N=24$, we show that $t_{\text{cusp}}\propto N$. Thus, for infinite system with $N\rightarrow \infty$, $t_{\text{cusp}}\rightarrow\infty$, and the obvious oscillation can not occur with finite time in the infinite system. We believe the finite-size analysis of the no thermalization regime is consistent with the MPS results shown in Ref. [29] and [30].

\end{document}